# Exploiting temporal network structures of human interaction to effectively immunize populations

Sungmin Lee[1], Luis E. C. Rocha[1], Fredrik Liljeros[2] and Petter Holme[1][3]

[1] *IceLab, Department of Physics, Umeå University, 90187 Umeå, Sweden*
[2] *Department of Sociology, Stockholm University, 10691 Stockholm, Sweden*
[3] *Department of Energy Science, Sungkyunkwan University, Suwon 440–746, Korea*

If we can lower the number of people needed to vaccinate for a community to be immune against contagious diseases, we can save resources and life. A key to reach such a lower threshold of immunization is to find and vaccinate people who, through their behavior, are more likely to become infected and effective to spread the disease than the average. Fortunately, the very behavior that makes these people important to vaccinate can help us finding them. People you have met recently are more likely to be socially active and thus central in the contact pattern, and important to vaccinate. We propose two immunization schemes exploiting temporal contact patterns. Both of these rely only on obtainable, local information and could implemented in practice. We show that these schemes outperform benchmark protocols in four real data sets under various epidemic scenarios. The data sets are dynamic, which enables us to make more realistic evaluations than other studies—we use information only about the past to perform the vaccination and the future to simulate disease outbreaks. We also use models to elucidate the mechanisms behind how the temporal structures make our immunization protocols efficient.

## Introduction

Contagious diseases that can be prevented by vaccination are a major burden to both rich and poor societies. They are estimated to account for an annual loss of 160 million disability-adjusted life years (DALYs)—around 10% of all diseases, or twice as much as HIV [1]. Vaccination of an entire community is often not possible due to limited supply and production capacity, especially for outbreaks of new disease. But to vaccinate a whole community is not desirable either—vaccine is expensive, it may have side effects and, luckily, it is not needed to immunize a community. If a large enough fraction $f$ of a community is vaccinated, a disease cannot spread to any substantial amount of people—the community has achieved *herd immunity* [2]. Lowering the threshold of $f$ to reach herd immunity is thus an important societal challenge that would help us saving resources, DALYs and life.

The epidemics of an infective disease is a complex function of both characteristics of the pathogen and the movement and interaction patterns of the people [2]. The diversity in people's contact patterns carries over to their importance for disease spreading. An outbreak such as the SARS epidemics of 2003 is thought not to have become a major event if it was not for a few highly influential spreaders [3] with a behavior far from average. To lower the threshold for herd immunity, it is crucial to identify and vaccinate these potentially influential individuals. The idea in this paper is to use the same contact structure that disease could spread over to identify important people to vaccinate. One early example of this idea is the *neighborhood vaccination* (NV) [4] protocol—choose person at random, ask her to name someone she met and vaccinate this other person. Chances are high that this other person has a large degree (number of neighbors) in the static interaction network and is influential in the disease spreading. The contact structure is thus not only influencing the disease dynamics, it is also a source of information that can be exploited to stop the disease. Now human interaction patterns have much more structure to be utilized in immunization protocols than just the distribution of degrees in a static network, what the neighborhood vaccination builds on. There is a great deal of dynamic structure too. The simplest dynamic contact patterns are cyclic—we are more likely to meet others 3PM than 3AM. Another temporal pattern is that there can be a broad distribution of contact rates between pairs of individuals. Especially for diseases that are not very contagious, this could give an impact on the disease dynamics that is hard to predict from the network structure alone [5]. A straightforward extension of the NV protocol to capture this structure would be to ask the person chosen at random to name the person she has met most often since some specific time. This is one of the protocols we test. A third temporal pattern, that static network models do not capture, is the tendency for two people to be in frequent contact for a period of time, followed by periods without meeting. This type of "bursty" [6] contact dynamics could also describe one person's behavior. In that case, it is important to vaccinate people who are in a period of activity, or will be in a near future. This leads to another extension of the NV protocol—ask the individual picked at random who her most recent contact (that could spread the disease) was, then vaccinate that person. Just like the NV protocol does this scheme not require any global knowledge and can be implemented in practice.

To briefly review subsequent developments, following the NV scheme, one line of research has focused on exploiting higher-order static network structure [7,8,9,10]. This type of immunization protocols has the obvious problem that higher-order structure is even more elusive to extract from social systems than the degree sequence of the contact network, and has, perhaps, its greatest applicability in stopping outbreaks of computer viruses. Another recent theme addresses the game-theoretic aspect of voluntary vaccination [11]. If the majority of a community gets vaccinated, the community has herd immunity and there is no need for a yet unvaccinated to get a shot. On the other hand, if few people get vaccinated, the risk of getting the disease grows and an injection may seem reasonable even to trypanophobics. The present work applies to scenarios of voluntary vaccination too, provided there is no strong correlation between an individual's behavior and her willingness to get vaccinated if faced with an approaching epidemics. Yet other studies of community immunization in a network-epidemiological framework focus on the simultaneous effects of the population's response to the disease and a vaccination campaign [12,13].

In the next section, we will present simulation results based on empirical contact patterns, comparing our proposed strategies to the NV scheme. Further on, we use a model generating correlated contact patterns to explain the different success of the schemes.

## Results

### The protocols

The two protocols we present in this paper use, like the neighborhood vaccination protocol, information from a random person $I$ in the community to find another individual to vaccinate that is more important to the disease spreading than $I$. The strategies are illustrated in Fig. 1.

In our first protocol, *Recent*, we iteratively ask a random individual $I$ to name the most recent contact (in a way that could transmit the disease in question) and vaccinate this person. The contact dynamics between two individuals has, at least in some circumstances, been observed to have a bursty dynamics—with alternating periods of activity and idleness [6]. The same pattern holds for the activity of individuals in the datasets we study in this work. The Recent protocol targets this type of temporal structure, and vaccinates individuals with a bias

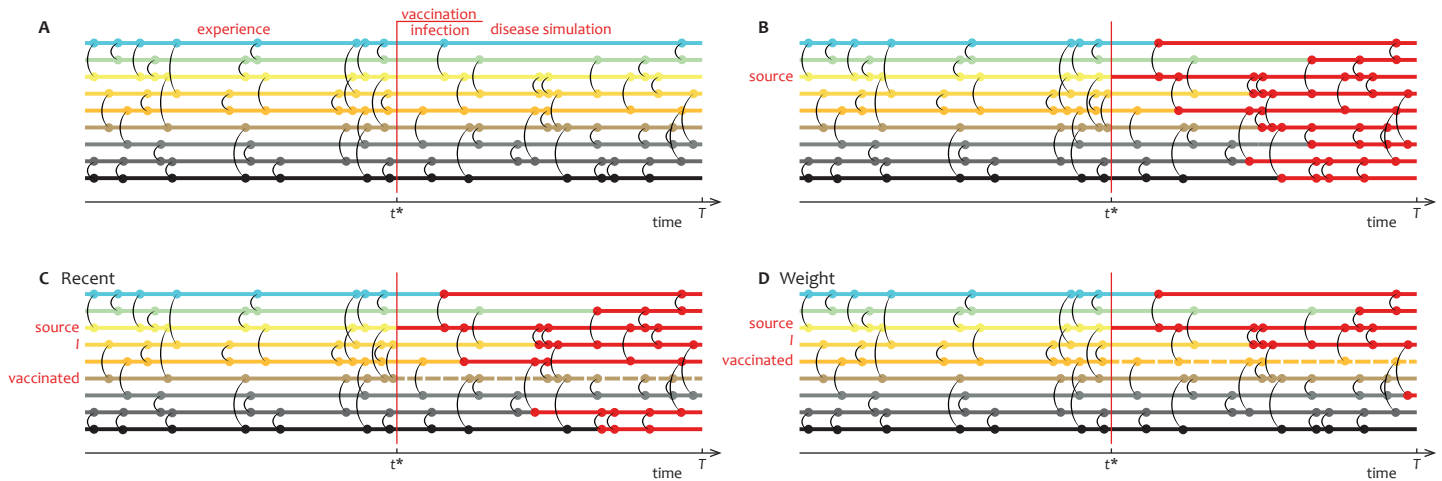

*Fig. 1.* An illustration of the simulations and the immunization protocols. Panel A displays an artificial contact structure where each horizontal line represents an individual. The circles and horizontal lines indicate the contacts. At a certain time (half the sampling time in this illustration, three quarters of the sampling time in the simulation), we both infect a random vertex and vaccinate a fraction f of the population, following a protocol. Panel B shows an example of a spreading process with 100% chance of contagion per contact and no recovery. Red lines represent infected individuals. In C and D, we see the same spreading event as in B in case one individual is vaccinated by the Recent (C) and Weight (D) strategies. I indicates the information-source vertex selected at random in the immunization scheme.

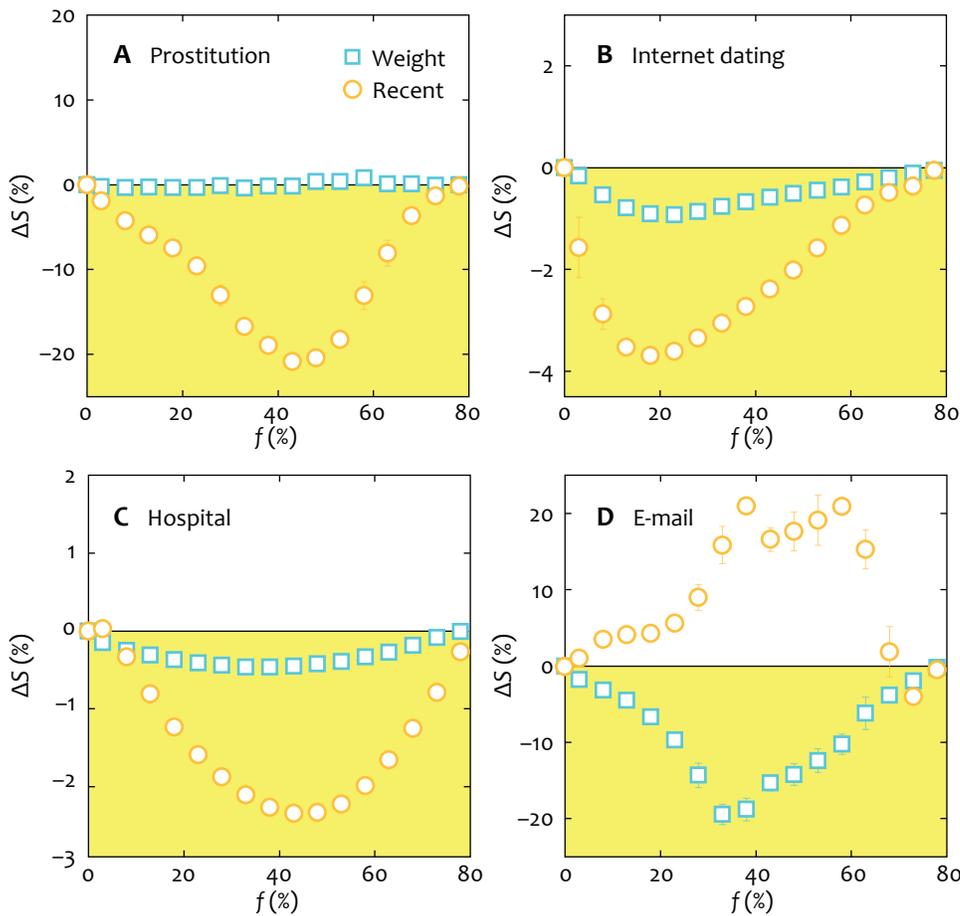

*Fig. 2.* The performance of the Recent and Weight strategies relative to the NV scheme. The performance measure S is the largest possible outbreak size, given the temporal contact structures, averaged over all infection sources. The yellow regions indicate an improvement on NV (the more negative values, the better). The different panels correspond to the four different data sets. The errorbars indicate standard errors over the set of infection sources.

towards those currently in a period of heightened activity.

In our second protocol, *Weight*, we iteratively ask a random individual $I$ to name their most frequent contact since some time $t$ before the present. This method seeks to vaccinate people that are, in general (or rather, over a longer timescale), more active than the average. To make the simulations more realistic, taking into account that people are not so accurate estimating the number of contacts they had in the past [14], we model this by assuming neighbors are vaccinated with a probability proportional to the weight of their tie with $I$ (so, the neighbor with highest weight would not necessarily be vaccinated).

**Empirical datasets**

We evaluate our strategies using four empirical datasets, more or less representative of the underlying structures for disease transmission. Two of the datasets come from online communication—one is the e-mail exchange data set from Ref. [6] where 3,188 email accounts were sampled over 83 days. An e-mail between two addresses in this set is recorded as a contact. In total there are 309,125 contacts. Emails to or from someone outside of the sampled email accounts are ignored. We do not claim this is a representative structure for disease spreading—information or opinions spreading is probably a more appropriate dynamic system in this case. We keep it as an example of a correlated, temporal network produced by humans and perhaps having some general features of human dynamics.

Our other communication data set comes from an Internet dating community [15]. Here various forms of communication between 29,341 members were recorded over 512 days amounting to 536,276 contacts. This data set is more interesting from an epidemiological viewpoint, both because the contacts are precursors of romantic and sexual relationship (and thus disease spreading), and that the community is closed so there is no bias from ignoring messages going in and out of the system.

In addition to the communication data, we use two data sets more directly linked to actual contact patterns. The first of these more real networks comes from a Brazilian forum where sex-buyers report and evaluate sexual encounters with escorts (top-end prostitutes). This data set spans 2,232 days, 16,730 people and 50,632 sexual contacts [16]. This data is of only a small fraction of the larger web of sexual contacts, which a full scale model has to take into account [17], but qualitative conclusions (affected by the type of temporal and topological correlations present, not their magnitude) should be valid even if we use this data as a raw contact structure.

Our fourth dataset records the proximity between patients in a hospital network. The data, described in detail in Ref. [18], covers 8,521 days and 295,107 patients of the Stockholm region of Sweden. If two patients are at the same ward at the same day we record that as a contact. In total there are 64,625,283 such contacts that can be interpreted as potential spreading events of nosocomial disease [19].

**Simulation of vaccination campaigns on empirical contact sequences**

The contacts of a population has two functions in a vaccination campaign. First, it is the basis of information on which we decide whom to vaccinate. Second, it is the connection structure actually spreading the pathogen. In our simulations, for the sake of serendipity, we divide the sampling timeframe $[0,T]$ into two periods $[0,t^*]$ and $[t^*,T]$ (where we chose $t^*$ as the time three quarters of the contacts have occurred) and use the first period as the information source for the immunization scheme, and the second for evaluating via disease simulation. This means that, in our study, the immunization program is assumed to occur at a time scale much shorter than the epidemics, which is strictly speaking not the case in reality. Our motivate for this assumption is that the results would probably be qualitatively the same without it, so to avoid the complication of scanning different vaccination rates, we assume the rate is infinite. We note also that vaccine is usually distributed in batches that make the vaccination process pulse-like rather than continuous. Another major assumption is that the disease is introduced in the system at the same time the vaccination program starts. While this is probably always, strictly speaking, incorrect, it is feasible to assume the vast majority of the population is uninfected at the time of the vaccination (otherwise other control measures, like travel restrictions, would probably be more appropriate [20]). A third unrealistic but simplifying assumption is that the vaccination is immediate and completely effective. Like the above assumptions, we make this to keep the model mathematically simple and similar to the rest of the literature. A more realistic model, with a non-zero probability of infection even though one is vaccinated, could be a topic for a deeper investigation, but would probably give similar results to a rescaling of $f$ (to smaller values, reflecting the occasional infection of a vaccinee).

**Largest outbreak sizes after vaccination**

Turing to the evaluation of the models, in Fig. 2, we plot the performance of the strategies as a function of the fraction $f$ of the population vaccinated. The performance measure is based on calculating $S$—the largest possible number of individuals that are infected at some point during an outbreak from any single source of infection over the set of contacts in the interval $[t^*,T]$. $S$ is thus a measure for dynamic contact-sequence data corresponding to the largest connected component (a common estimate of the severity of worst case scenarios [21]) in static networks. Note that $S$ goes beyond pure network modeling [17,22,23] and captures some effects of the temporal ordering of contacts (such as that A cannot infect C via B if all contacts between B and C occurs before all contacts between A and B). To emphasize the relative benefits of the different strategies, we plot the fractional increase of S with respect to the NV scheme, ΔS. If, for example, ΔS = −10% our strategy in question decrease the maximum outbreak size 10% relative to neighborhood vaccination.

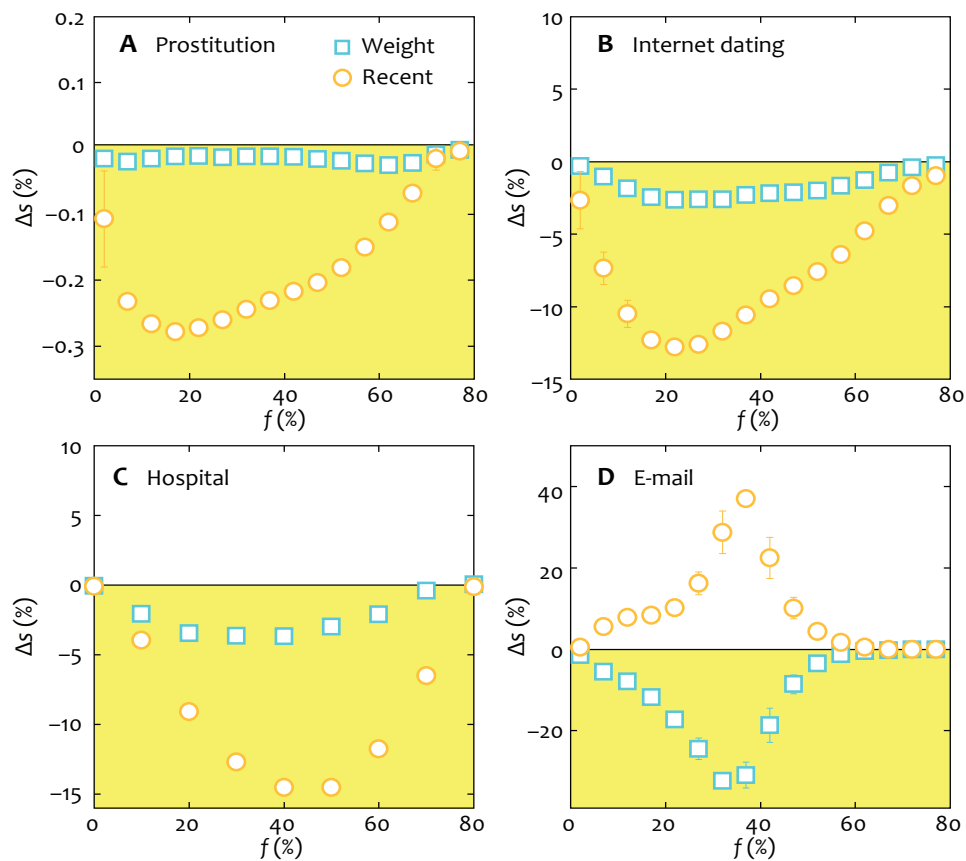

*Fig. 3.* The performance of the Recent and Weight strategies relative to the NV scheme for a dynamic, SIS-type disease simulation. The performance measure in this case is the average outbreak size $s$ (total number of infected individuals) in a SIS simulation with a per-contact transmission probability $\lambda = 0.25$ and a duration $d$ of the infected stage of three weeks. Just like Fig. 2, the vaccination is more efficient, relative to NV, the lower $\Delta s$ is. The errorbars correspond to the standard error calculated over all unvaccinated vertices as infection sources and 1000 runs of the vaccination and SIS simulation per source.

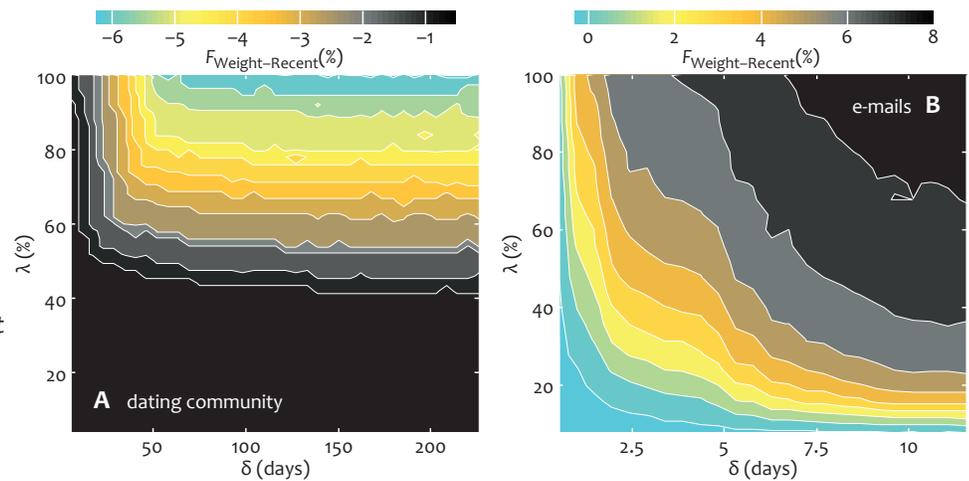

**Fig. 4.** Contour plots showing the relative performance of Weight and Recent over a range of parameter values of the SIS model. We measure the fraction of nodes i such that if an outbreak started at i it would have been most efficiently stopped by an immunization campaign by the Weight strategy (as opposed to Recent). To let 0 represent neutrality, we subtract 1/2 from the above-mentioned fraction to achieve at $F_{Weight-Recent}$, the plotted performance measure. For every parameter value, we use all unvaccinated vertices as infection sources and 100 runs of the immunization protocol and disease simulations. In this plot we use f = 20%. Our other data sets (from prostitution and hospital contacts) behave qualitatively like the dating-community data (A).

Our first conclusion is that ΔS is mostly negative—both Recent and Weight outperform NV for most datasets and fractions of the population vaccinated. Weight is in almost never worse than NV. Recent, on the other hand, performs worse than NV for the email data but is, for the other contact sequences, a clear winner. The prostitution, Internet dating and hospital data all gives similar results for ΔS, the curves for the e-mail data look drastically different. These two distinct behaviors are something we will se for other quantities as well. The relative advantage of Recent is strongest for the sexual contact data of Fig. 2A—this is also true for our other analyses.

**Average outbreak sizes in dynamic simulations**

Although S is a very simple measure, the maximal outbreak size depend on temporal contact structures. However, there are other synergetic effects between the contagion and contact dynamics that it does not capture. If, for example, there is an anticorrelation between the accumulated number of contacts and degree (as has been reported for sexual networks [5]), then the outbreak distribution would likely be more narrow (as there it has a significant chance of dying early even though it starts at a high-degree vertex) than if there is a strong positive correlation between activity (the total number of contacts in the data) and degree, as is the case in the email network. To test the immunization protocols in a more realistic situation than the worst-case scenarios (of maximal per-contact transmission probability λ), we also run Susceptible–Infected–Susceptible (SIS) simulations. In our simulations a susceptible individual become infected upon contact with an infected with a probability λ. We let the infected stage last a fixed duration δ. We go through all unvaccinated nodes as sources of infection and simulate the disease spreading in the interval [t*,T]. It might thus happen that the initially infected individual is only present in the data before t*, in which case the outbreak size will inevitably be one.

The first quantity we look at for these simulations, see Fig. 3 (shows results for SIS), is the average outbreak size s (averaged over all unvaccinated individuals as infection sources and 1000 random seeds) as a function of f. For this plot we use the parameter values λ = 0.25 and δ = 3 weeks. We choose these parameter values reflect intermediately contagious diseases (less than Syphilis, more than HIV) and short durations to capture dynamic effects of the finite duration of diseases. Since the data sets are limited in time, such effects would vanish if δ was much longer. The SIR (Susceptible–Infected–Removed) model with the same parameter values gives rather similar curves—the skewed distribution of activity in these data sets means that the probability of reinfection (the difference between SIS and SIR) is significant only for the comparatively small group of the most active individuals.

The curves from Fig. 3 are strikingly similar to Fig 2. Only the magnitude of the differences varies—for the prostitution Δs (Fig. 3A) is consistently smaller than ΔS (Fig. 2A); for the other three data sets, the difference in performance is larger for the SIS simulations compared to the worst-case scenario measure. One explanation for the small differences in the prostitution data is that about three quarters of the contacts occur only once. Our strategy Recent can eliminate a worst case scenario by finding people involved in these, rather rare, recurring contacts, but for average outbreak sizes measured in the SIS simulations, the chance of an outbreak is so small that the *s*-values do not differ much.

**Relative advantage of strategies as a function of infectivity and duration of the infective state**

We continue our analysis of how the vaccination affects the average outbreak sizes in stochastic simulations by looking at the response of *s* to the model parameters λ and δ. In this analysis, we keep *f* = 20%—a value close to where the choice of immunization strategy makes most difference. Also in this analysis, the four datasets fall into two classes where the email data has a unique behavior and the three others are similar to each other. To get better converged plots, we let the smallest dataset, the Internet dating data, represent this class of similar behavior (also comprising the prostitution and hospital datasets). To evaluate the strategies, we go through all the unvaccinated individuals as initial sources of the epidemics, apply the immunization protocols, and calculate for a pair A and B of immunization strategies: out of 100 runs of the SIS model how many times does strategy A outperform strategy B. In Fig. 4 we present the deviation in percent $F_{Weight-Recent}$ from a scenario where the strategies are equally successful. The main conclusion is that the observation from the Fig. 3 holds throughout the (λ,δ) parameter space—Weight is the best strategy for the email data, Recent is the best for the others. In the small λ and small δ limit the disease will die out soon whether someone has been vaccinated or not. This explains why the smallest deviations, both in Fig. 4A and B, occurs for the smallest (λ,δ)-values. Then, if we focus on the dating community plot Fig. 4A for a moment, there is a dramatic change in $F_{Weight-Recent}$ as λ reaches above about 50% for δ > 40 days. This is related to a epidemic threshold that, despite the skewed degree distributions, is rather conspicuous for this type of data [17]—for λ > 50% a disease can spread to a finite fraction of the population and the immunization protocols do make a difference for this data set. Furthermore, if one varies δ, *F* respond in a highly non-linear—if the duration of the infection is long enough, the benefits of the strategies are similar, but for diseases short in duration, *F* changes rapidly with δ. For the e-mail data there is a similar plateauing δ-dependence of *F*, but the λ-dependence more dramatic close to zero, rather than an intermediate value.

**Model of artificial contact sequences**

From the above studies we can conclude that, regardless of the type of the disease, Recent is the best strategy for the prostitution, Internet dating and hospital proximity data, whereas Weight is the better strategy for the e-mail dataset. Why? One aspect that sets aside the e-mail data is that it is the user behavior seems fairly stationary (48% of the users are present in both the first and last 5% of the contacts). In such a situation where the overall activity is rather uniform, the activity of the more distant past could predict the future activity.

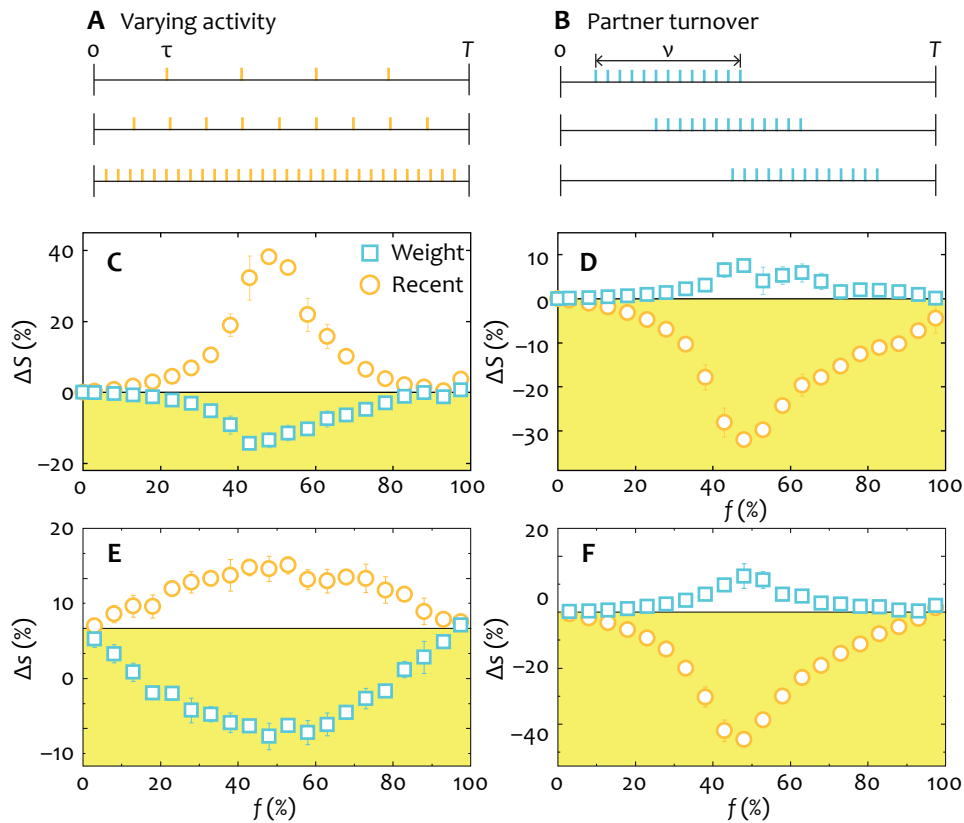

*Fig. 5. Evaluating the performance of the vaccination strategies for different types of temporal correlations. In A and B we illustrate the models that encode the different temporal contact structures. In the varying activity model (A) the first contact along an edge happens at time $t_s$ after the beginning of the simulation and then subsequent contacts happen with a time interval $t_s$. In the other, partner turnover, model (B), an edge becomes active with uniform probability in time the interval $[0, T – v]$. The edge is active for v time steps with one contact per time step.*

The Internet dating, hospital and prostitution data are more non-stationary with individuals both entering and leaving the system (here, only 1.0%, 0.3% and 0.8% individuals, for the three data sets, respectively, are present both in both the first and last 5% of the contacts). The Internet dating community has an increasing trend in the system-wide activity level, whereas the hospital and prostitution data shows more of a quasi-stable behavior where individuals enters and leaves the system at a more constant pace. If we assume a situation where activity-wise identical users comes and leaves the system at equal rates, the users most recently seen active are also the ones most likely to be active in the near future, and thus the ones most urgent to vaccinate. Therefore we can understand Recent is the best strategy for the Internet dating, prostitution and hospital proximity data.

To put the arguments above on a more solid footing, we construct two models of contact patterns capturing these two temporal structures see illustrations in Figs. 5*A* and *B*. In both these models the network structure is purely random (details in the Methods section) to ensure that all effects we observe are temporal. In the first model, capturing a *varying activity* (the *VA* model), we let the communication over an edge (a connected pair of vertices in the network) take place at intervals of $\tau$, a value drawn from a uniform distribution, until time reaches *T*. The second model embodies the birth and death of relationships—every edge is active for a fixed duration ($\Delta t$ time steps, with one contact per time step), but the starting time is random. We call this model the *partner turnover* (*PT*) model.

In Figs. 5*C–F* we plot the results from our simulations of the contact pattern models. These simulations, investigating both worst-case scenarios (Figs. 5*C–D*) and average outbreak sizes in the SIS model (Figs. 5*E–F*), confirm that temporal structure can create the different efficacies of the immunization protocols. For the VA model, since the neighbor to vaccinate is chosen in proportion to the weight, the chance of picking a highly active individual is higher with the Weight strategy than NV. If Recent is applied to the VA model in our range of parameters (relatively large $t*$) there is a heightened chance the latest contact is one with a small $\tau$ that will not happen again, which makes Recent perform worse than NV. For the PT model, if an edge recently had some activity, chances are high that it will be active soon again. Recent is designed to find such recently active edges, so it logically works better than NV in this situation. If there are relationships that are over in the PT model, then Weight will pick one of those. This is clearly counterproductive compared to, like the NV strategy, just choosing a random neighbor. One can show, using the accumulated degree as a proxy for the importance of the vaccinated vertex, that Recent performs better than NV that performs better than Weight for the partner turnover model and that Weight performs better than NV that performs better than Recent for the varying activity model, for most realistic parameter values.

## Discussion

The real contact structures over which diseases spread are structured not only in network topology but also temporally. Just like the network structure can affect the dynamics of the disease spreading, so can the temporal contact structure; and just like vaccination strategies can exploit the topology, they can also exploit the temporal structure. In this paper we propose two such immunization protocols that are practically realizable, utilize both temporal and topological network structure, and are more effective than the neighborhood vaccination protocol (that only uses static topological information). The two strategies are based on sampling persons at random, asking them "who where you in contact with most recently [in such a way that disease could have spread]?" (the Recent strategy) or "who were you in contact with most often the last *X* months?" (the Weight strategy), and vaccinate the named person. In this paper we study the long time limit of *X*, so that the *X* months cover all the data set before the vaccination time.

We test the strategies on four empirical, dynamic datasets of contacts. Two of these datasets represent possible pathways for real epidemics (a sexual network of Internet-mediated prostitution and a proximity network of patients in a hospital system). The other two datasets come from online communication—one representing interaction at an Internet dating site (where an unknown fraction of the contacts probably turn into off-line contacts with a possibility of contagion), the other is an email dataset (where, presumably, an edge means a high chance of an offline social tie, but the temporal contact structure is probably not so correlated with the offline contacts). We split the sampling times of the data into two parts, the first where the individuals experience the world, the second where an epidemics spread over the contacts. In contrast to other vaccination stimulations [4,7,8,9,10], we do not assume the contact patterns are the same before and after the vaccination. In these other works, the network that will transmit the disease after the vaccination is already used as a basis to identify the individuals to vaccinate. In this respect, our approach is more strict and realistic compared to the abovementioned studies.

As it turns out, the Weight strategy outperforms Recent and NV for the e-mail data while Recent is the most efficient scheme for the other three datasets. This tells us four things: first, that there is enough temporal structure in the contact patterns for our protocols to be effective; second, that the optimal choice of immunization protocol can be dependent on the specific contact structure of a disease; third, that in the more realistic networks Recent is the better strategy; and fourth, that the temporal correlations of these more realistic network are relatively short. After a closer look at the temporal structures separating these data sets, using models of contact dynamics (that tunes the temporal structure, but not the topological), we argue that a turnover of relationships promotes the efficiency of Recent. Weight, on the other hand, is most efficient when the ties between individuals are strongly overlapping in time, but there is a broad distribution of contact rates over those ties. In sum, with the information we have and no way of making a closer estimate of the contact structure, we would recommend vaccinations according to the Recent scheme. In most countries, most vaccination campaigns are voluntary, and even if they were not, they will necessarily contain errors. On the other hand, in the limit of many errors the protocols will all approach a random vaccination (which is probably, in practice, close to voluntary vaccination). Any step away from this limit would mean that more information from the contact patterns reach the immunization scheme which would mean give an improvement on the NV scheme.

To get a little more concrete assessment of the effects of the inaccuracy of interviewees, one can compare our immunization protocols with respondent-driven sampling (RSD)—a method for sampling hidden populations that navigates between people by questions in a way similar to the protocols [24]. Respondent-driven sampling rests on stronger assumptions about what people know about their social surroundings, and is statistically more venturesome, than the problem of finding efficient immunization protocols. However, empirical studies shows that, even if RDS needs larger samples than it generates to estimate variances, it does converge in means for populations of hundreds [25]. Since the immunization problem does not stuffer from the uncontrollable sample sizes of the snowball-sampling based RDS, we take Ref. [25] as an indication that our methods would be efficient in practice.

On a fundamental level, the fact that temporal structures alone can influence the efficiency of immunization protocols not less than the topology needs to be more thoroughly understood. Similar problems arise in other areas where one want to limit a spreading phenomenon dependent on the contact dynamics, like disease in wild or domestic animals, e-mail viruses or computer worms [26]. We anticipate more research in this direction.

## Methods

**Disease-spreading simulation**

All the datasets we use can, mathematically, be represented as lists of *contacts* $(x_i, y_i, t_i)$, $i = 1, \ldots, C$. Each triple represents a contact between individual $x_i$ and $y_i$ at time $t_i$. We can assume that a contact list is ordered so that $t_i \leq t_{i+1}$ for all $i$. Without loss of generality, we set $t_1 = 0$. $T$, the total sampling time, is thus simply $t_C$. Let $N(t)$ be the number of vertices at time $t$ and $N$ (without argument) denote $N(T)$.

We divide the sampling time into two parts $[0, t^*]$ and $[t^*, T]$ where $t^* = t_{3C/4}$. At time $t^*$ we both vaccinate the population and start the disease. We choose one vertex among the entire unvaccinated population (even if their last contact is before, or first contact after, $t^*$), with uniform randomness, as an infection source. The immunization protocols use the experience from the interval $[0, t^*]$, but no information whatsoever about the interval when the epidemics is unfolding $[t^*, T]$.

At time $t^*$ we choose $Nf$ individuals to vaccinate (where $f$ is a control parameter setting the fraction of the population to vaccinate). The flow chart of the simulation is:

1. With uniform probability, pick an individual $i$ among the $N(t^*)$ individuals present in the data at this time.
2. Pick a neighbor $j$ of $i$, either the most recent contact of $i$ (the Recent protocol) or the most frequent contact in the interval $[t^* - X, t^*]$, $0 \leq X < t^*$ (Weight). For simplicity, we use $X = t^*$ in this paper.
3. If such a vertex $j$ exists and is not vaccinated, then vaccinate $j$.
4. If yet not $Nf$ vertices are vaccinated, go to step 1.

One run of the SIS disease simulation starts by one source vertex being marked infected, and all other vertices marked susceptible. Then we go through all contacts $(x_i, y_i, t_i)$, $3C/4 < i \leq C$, and if $x_i$ ($y_i$) is infected, but not $y_i$ ($x_i$), then, with a probability $\lambda$, $y_i$ ($x_i$) becomes infected. After a time $\delta$ an infected vertex becomes susceptible again. Our key quantity is $s$—the total number of infected vertices at time $T$. When we study the average largest possible outbreak size—technically equal to the outcome of a SIS simulation with $\lambda = 1$ and $\delta = \infty$—we use the symbol $S$ (instead of $s$) for the average number of individuals that can be reached by successive contacts from the source. To calculate $s$ and $S$, we average over all $(1 - f)N$ unvaccinated vertices as infection sources and 1000 independent runs of the immunization protocol and disease propagation.

**Models of contact dynamics**

To elucidate the effects of the temporal structure on the immunization protocols, we use two generative models of contact sequences. The network topologies of these simulations are the same—an instance of an Erdős–Rényi model [27] 1000 vertices and 2000 edges. Given the topology, we associate every edge with a set of contacts generated by one of two schemes. For the first scheme (the varying activity model), we draw a random number $\tau$ with uniform probability in the interval $[0, T]$. Then we let the contacts over the edge take place at times $\tau$, $2\tau, \ldots, n\tau$, where $n$ is the largest number such that $n\tau < T$. In the other scheme, the partner turnover model, the contacts take place over $\Delta t$ consecutive time steps. The starting time for this burst of contacts, $t_s$, is a random variable drawn with uniform probability from the interval $[0, T - \Delta t]$.